\documentclass{cernyrep}

\usepackage{amsmath,amssymb,physics}
\usepackage{siunitx}
\usepackage{graphicx}
\usepackage{float}

\usepackage[bookmarks, colorlinks=true, linktoc=page, pdftex, linkcolor=black, citecolor=black, urlcolor=blue]{hyperref}
\usepackage{tcolorbox}
\usepackage{enumitem}
\usepackage{booktabs}
\usepackage{tikz}
\usepackage{pgfplots}
\pgfplotsset{compat=1.18}
\usetikzlibrary{arrows.meta, positioning}

\usepackage{fancyhdr}
\usepackage{truncate}
\usepackage[a4paper,margin=2.5cm]{geometry}
\usepackage{xcolor}
\usepackage{caption}
\tcbset{definition/.style={colback=blue!4,colframe=blue!55!black,boxrule=0.6pt,arc=2pt,left=6pt,right=6pt,top=6pt,bottom=6pt}}
\setlist[itemize]{leftmargin=*,topsep=3pt,itemsep=2pt}
\usepackage{fancyhdr}
\fancyhfoffset{4 mm}
\fancypagestyle{ARTTITLE}{%
\fancyhf{}
\lhead{\raggedright {\it Mechanical \& Materials Engineering for Particle Accelerators and Detectors}\raggedright \\ CERN Accelerator School Proceedings ---  Sint-Michielsgestel,  Netherlands, 2024\hfill}
\lfoot{\hspace{3mm} Available online at \url{https://cas.web.cern.ch/previous-schools}}
\rfoot{\thepage\hspace*{3mm}}

}

\usepackage{graphicx}
\usepackage{booktabs}
\usepackage{siunitx}
\usepackage{float}

\begin{document}
\title{An Introduction to Measurement Uncertainty}
\author{Samanta Piano
\institute{Manufacturing Metrology Team, Faculty of Engineering,\\ University of Nottingham, Nottingham, UK}}

\begin{abstract}
This chapter introduces the fundamental principles of metrology and the concept of measurement uncertainty. It explains the role of measurement in engineering and manufacturing, outlines the distinction between error and uncertainty, and presents standard methods for evaluating uncertainty, including the GUM framework, uncertainty budgets, and Monte Carlo simulation. Practical examples and industrial standards are discussed to illustrate real-world applications.
\end{abstract}

\maketitle
\thispagestyle{ARTTITLE}

\section{Introduction to metrology}

\begin{tcolorbox}[definition]
Metrology is the science of measurement and its application. It encompasses both the theoretical principles underlying measurement and the practical techniques used to obtain reliable quantitative information about physical quantities. It also includes the definition of measurement units, the establishment of standards, and the methods used to ensure accuracy, consistency, and traceability.
\end{tcolorbox}
This definition follows the terminology adopted in the \textit{International
Vocabulary of Metrology (VIM)} \cite{VIM2012}, which provides internationally
agreed definitions for the fundamental concepts used in measurement science.
A key role of metrology is to enable comparability of measurements performed across different locations and times. This is achieved through internationally agreed standards and traceability chains, ensuring that measurements correspond to the same physical quantities globally. In practice, metrology provides the framework that allows measurements
performed in different places and at different times to be compared in a
consistent manner. Without such a framework it would be impossible to ensure
that a measurement performed in one laboratory corresponds to the same
physical quantity measured elsewhere. The establishment of common measurement
units and traceability chains ensures global comparability of measurement
results \cite{BIPM2019}.

Modern metrology is usually divided into three broad areas \cite{Leach2020}: 
\begin{itemize}
    \item \textbf{Scientific
metrology} focuses on the development of fundamental measurement standards and
the realisation of units within the International System of Units (SI).
\item \textbf{Industrial metrology} applies measurement science to manufacturing processes in
order to guarantee that products meet design specifications. 
\item \textbf{Legal metrology}
deals with measurements that have regulatory or commercial importance, such
as weighing systems used in trade or instruments used in medical diagnostics.
\end{itemize}

\subsection{Importance of measurements}
\label{sec:sss}
Measurement plays a central role in almost every field of modern science and
technology. Scientific experiments rely on accurate measurements to test
theories and discover new physical phenomena \cite{taylor1997error}. Engineering design requires
precise measurements to verify that components meet their intended
specifications \cite{BIPM2019}. The reliability of these measurements directly affects the safety, quality
and efficiency of many technological systems. For example, in aerospace
engineering the dimensions of structural components must be verified with
extremely high precision to ensure correct mechanical behaviour \cite{leach2014optical}. In medical
applications, accurate measurements are essential for determining appropriate
drug dosages and monitoring physiological parameters. More generally, measurements provide the quantitative information required
for decision-making in science, engineering and industry \cite{GUM2008}. If measurements are
unreliable or poorly understood, the conclusions drawn from them may also be
incorrect.

\subsubsection{Measurements in manufacturing}
Manufacturing processes aim to produce components with specific geometric
dimensions and material properties defined during the design stage. These
specifications are typically expressed in terms of nominal values and
tolerances. In an ideal manufacturing process every component would match the nominal design value exactly. In reality, however, production processes inevitably generate variability. Factors such as tool wear, machine vibrations, temperature fluctuations and variations in material properties cause the dimensions of produced parts to deviate slightly from the intended value. As a consequence, the dimensions of manufactured parts generally follow a
statistical distribution around the nominal value rather than a single exact
value. Measurement systems are therefore essential in manufacturing for monitoring this variability and ensuring that products remain within acceptable tolerance
limits. Accurate dimensional measurements allow engineers to determine whether
components satisfy the required specifications and to detect deviations in the production process \cite{Leach2020}. Modern manufacturing increasingly relies on advanced measurement systems such as coordinate measuring machines, optical interferometers and structured
light systems. These technologies enable high-precision measurements that are essential for industries such as aerospace, semiconductor manufacturing and precision engineering.
\subsection{From measurement to measurement uncertainty}
An important concept that arises in metrology is that no measurement is
perfectly exact. Every measurement result contains some degree of uncertainty
arising from instrument limitations, environmental conditions and measurement
procedures.

The systematic evaluation of this uncertainty is therefore an essential part
of modern measurement science. The internationally accepted framework for
evaluating measurement uncertainty is provided by the \textit{Guide to the
Expression of Uncertainty in Measurement (GUM)} \cite{GUM2008}, which will be
discussed in detail in the next sections.

\section{Measurement uncertainty}

In the previous section we discussed the importance of measurement in
engineering and manufacturing. An important idea that arises naturally
from that discussion is that no measurement can be perfectly exact.
Every measurement result contains some degree of doubt arising from
instrument limitations, environmental influences and the measurement
procedure itself. The purpose of measurement uncertainty analysis is to quantify this
doubt. Instead of reporting a measurement result as a single number,
modern metrology expresses the result together with an interval that
indicates how reliable that number is. Understanding and evaluating measurement uncertainty is therefore a
central task in metrology and quality control.

\subsection{Definition of measurement uncertainty}

The internationally accepted definition of measurement uncertainty is
provided in the \textit{International Vocabulary of Metrology (VIM)}
\cite{VIM2012}.

\begin{tcolorbox}[definition]
Measurement uncertainty is a non-negative parameter that characterises
the dispersion of the quantity values that could reasonably be
attributed to a measurand.
\end{tcolorbox}

In simpler terms, measurement uncertainty represents the range of values
within which the true value of the measured quantity is believed to lie. The concept can be illustrated by considering repeated measurements of
the same quantity. Even if the measurement conditions remain unchanged,
the results will generally vary slightly. This variation reflects the
combined influence of multiple uncertainty sources. Rather than attempting to determine the exact true value, metrology
focuses on estimating the~interval that is most likely to contain that
value.

\subsubsection{Measurement result and uncertainty}

A measurement result is normally expressed in the form
\begin{equation}
Y = y \pm U,
\end{equation}

where $y$ is the measured value and $U$ is the expanded uncertainty. The expanded uncertainty defines an interval around the measured value
that is expected to contain the true value with a specified level of
confidence. For example, a measurement reported as

\begin{equation}
L = (50.00 \pm 0.02) \text{ mm}
\end{equation}

indicates that the true length is expected to lie between
\SI{49.98}{mm} and \SI{50.02}{mm} with the chosen confidence level.

\subsubsection{Error and uncertainty}

It is important to distinguish between the concepts of error and
uncertainty \cite{VIM2012,GUM2008}, which are often confused. The measurement error is defined as the difference between the measured
value and the true value:
\begin{equation}
\epsilon = x_{\mathrm{measured}} - x_{\mathrm{true}}.
\end{equation}

However, the true value is generally unknown. Because of this, the error
cannot usually be determined exactly. Measurement uncertainty therefore describes the possible magnitude of the error rather than the error itself.

\begin{tcolorbox}[definition]
Measurement error is the difference between a measured value and the
true value of the quantity. Measurement uncertainty quantifies the
possible range of that error.
\end{tcolorbox}
This distinction is fundamental in modern metrology. Instead of
attempting to eliminate all errors completely, the goal is to evaluate
and minimise their influence on the measurement result \cite{BIPM2019}.
Every measurement result differs, to some degree, from the true value of
the quantity being measured. The~difference between the measured value
and the true value is known as the measurement error. In practice, the~true value is rarely known exactly. As a result, the
error cannot usually be determined directly. Instead, metrology focuses
on understanding the mechanisms that generate measurement errors and on
estimating their possible magnitude through the concept of measurement
uncertainty. Measurement errors are generally classified into two main categories:
systematic errors and random errors.

\subsubsection{Systematic errors}

A systematic error is a consistent deviation of the measured values from
the true value. When systematic errors are present, repeated
measurements tend to be biased in the same direction.

Systematic errors typically originate from imperfections in the
measurement system or from environmental influences that affect the
measurement in a predictable way. For example, a miscalibrated
instrument may consistently report values that are slightly larger or
smaller than the true value. Similarly, temperature changes may cause
thermal expansion of mechanical components, introducing a predictable
bias in dimensional measurements. An important characteristic of systematic errors is that they cannot be
reduced simply by repeating the measurement. If the measurement system
contains a systematic bias, all repeated measurements will be affected
in a similar way. However, systematic errors can often be corrected once they are
identified. Calibration procedures are commonly used to detect and
compensate for systematic deviations in measurement instruments.

\subsubsection{Random errors}

Random errors arise from unpredictable fluctuations in the measurement
process. Unlike systematic errors, random errors cause measurement
results to vary in an irregular manner when the same quantity is
measured repeatedly. These fluctuations may originate from sources such as electronic noise
in sensors, environmental disturbances, vibration or small variations in
the measurement procedure. Because the effects of these influences vary
from one measurement to another, the resulting errors appear random. Random errors cannot be completely eliminated, but their influence can
be reduced by performing repeated measurements and analysing the
statistical distribution of the results. Averaging multiple measurements
typically improves the reliability of the estimated value.

\subsection{Systematic and random error illustration}

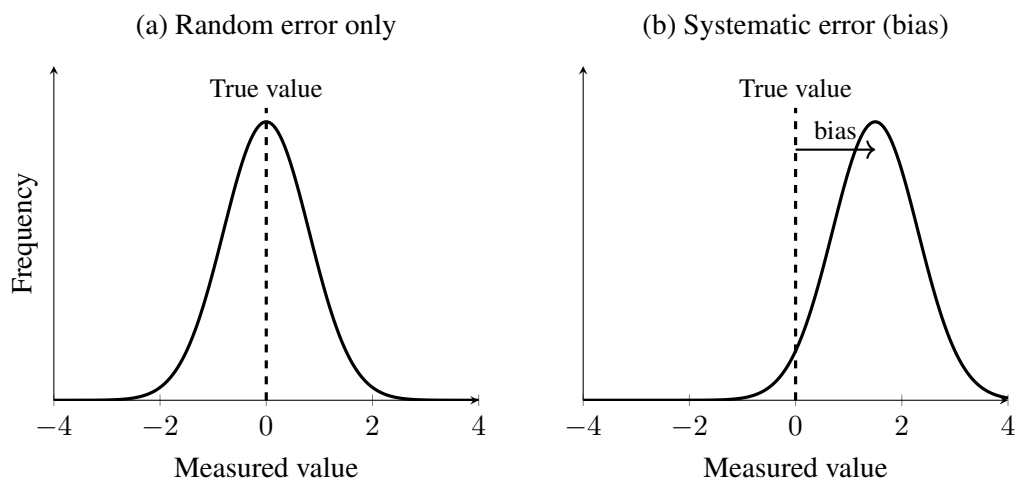
\begin{figure}[H]
\centering

\begin{tikzpicture}


\begin{axis}[
width=0.45\textwidth,
height=6cm,
xmin=-4,xmax=4,
ymin=0,ymax=1.2,
axis lines=left,
xlabel={Measured value},
ylabel={Frequency},
title={(a) Random error only},
ytick=\empty
]

\addplot[dashed,very thick] coordinates {(0,0) (0,1.05)};
\node[above] at (axis cs:0,1.05) {\small True value};

\addplot[very thick,domain=-4:4,samples=200]
{exp(-x^2/1.3)};

\end{axis}


\begin{axis}[
at={(7cm,0)},
width=0.45\textwidth,
height=6cm,
xmin=-4,xmax=4,
ymin=0,ymax=1.2,
axis lines=left,
xlabel={Measured value},
title={(b) Systematic error (bias)},
ytick=\empty
]

\addplot[dashed,very thick] coordinates {(0,0) (0,1.05)};
\node[above] at (axis cs:0,1.05) {\small True value};

\addplot[very thick,domain=-4:4,samples=200]
{exp(-(x-1.5)^2/1.3)};

\draw[->,thick] (axis cs:0,0.9) -- (axis cs:1.5,0.9);
\node[above] at (axis cs:0.75,0.9) {\small bias};

\end{axis}

\end{tikzpicture}

\caption{Illustration of measurement errors. 
(a) When only random errors are present, repeated measurements scatter 
around the true value. 
(b) When a systematic error exists, the distribution of measured values 
is shifted away from the true value, producing a bias.}

\label{fig:systematic_random_error}
\end{figure}

The difference between systematic and random errors can be illustrated
using repeated measurements of the same quantity, as shown in Fig. \ref{fig:systematic_random_error}. If only random errors are present, the measurement results will be
distributed around the true value, forming a statistical distribution
such as a normal distribution (Fig. \ref{fig:systematic_random_error} (a)). If a systematic error is present, the distribution of measured values
will be shifted away from the true value, indicating the presence of a
bias in the measurement system (Fig. \ref{fig:systematic_random_error} (b)).
Table \ref{tab:systematic_random} gives a summary of the comparison between systematic and random error.

\begin{table}[H]
\centering
\fontsize{10}{12}\selectfont
\caption{Comparison between systematic and random errors in measurements.}
\label{tab:systematic_random}

\begin{tabular}{p{6cm} p{6cm}}
\hline \hline
\textbf{Systematic Error} & \textbf{Random Error} \\
\midrule

Produces a consistent deviation from the true value (bias). &
Produces unpredictable variations around the true value. \\

Often caused by calibration errors, instrument drift,
misalignment or environmental effects. &
Often caused by noise, vibration, electronic fluctuations
or small variations in the measurement process. \\

Cannot be reduced by repeating the measurement. &
Can be reduced by averaging repeated measurements. \\

Often difficult to detect without calibration or comparison
with reference standards. &
Can be detected through statistical analysis of repeated measurements. \\

Can usually be corrected once identified (e.g. calibration). &
Cannot be completely eliminated, only statistically reduced. \\

Affects the \textbf{accuracy} of the measurement. &
Affects the \textbf{precision} or repeatability of the measurement. \\

\hline \hline
\end{tabular}

\end{table}

Understanding the distinction between systematic and random errors is
essential for evaluating measurement uncertainty. Random errors contribute directly to the statistical spread of
measurement results and are therefore evaluated through statistical
analysis. Systematic errors, on the other hand, must be identified and
corrected whenever possible. Any remaining uncertainty associated with
these corrections must then be included in the overall uncertainty
budget. Modern uncertainty evaluation methods therefore combine information
about both types of error in order to estimate the overall reliability
of a~measurement result.

\subsection{Sources of measurement uncertainty}

\begin{figure}[tb]
\centering
\includegraphics[width=0.95\textwidth]{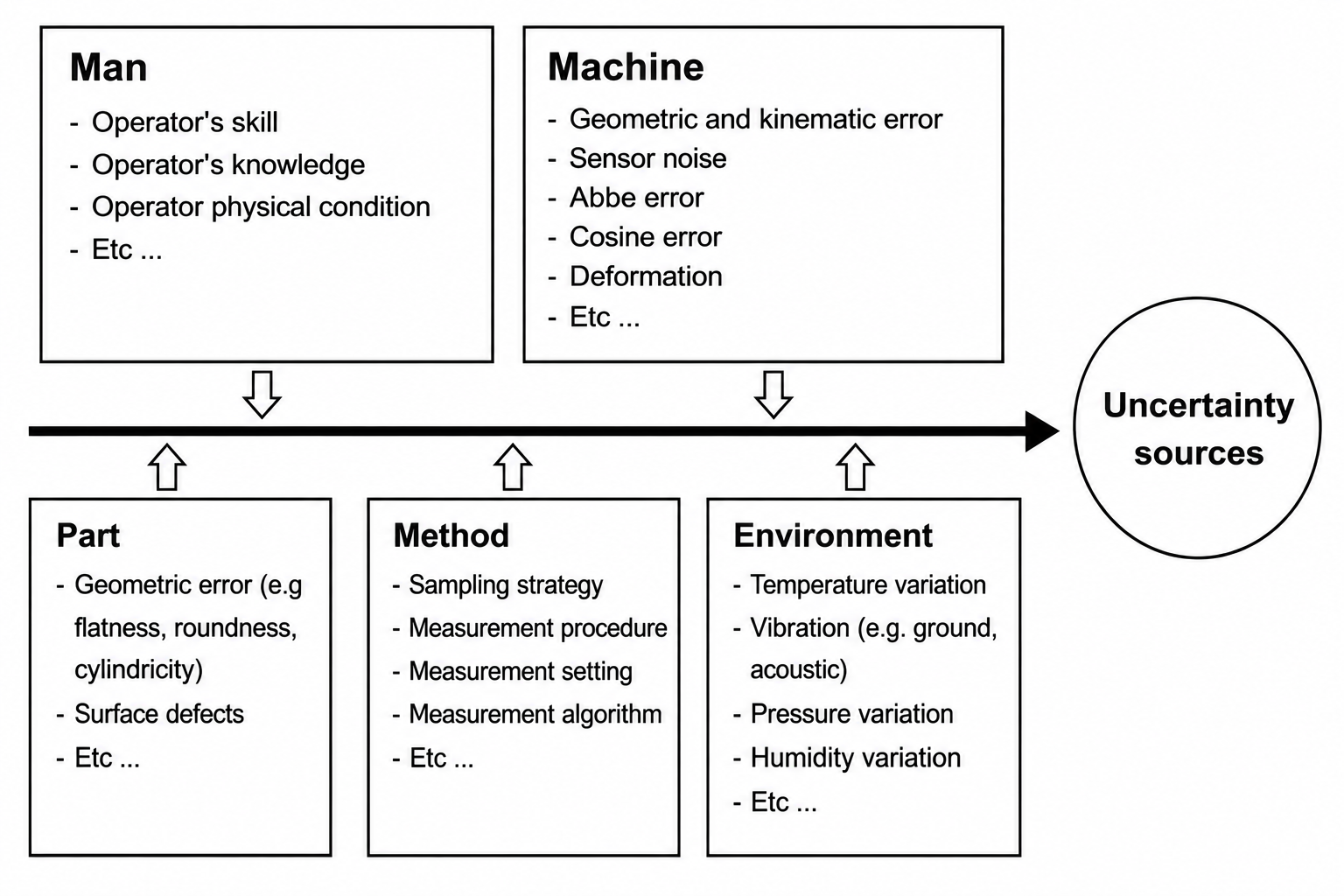}
\caption{Typical sources contributing to measurement uncertainty.}
\label{sources_of_uncertainty}
\end{figure}

Measurement uncertainty arises from many different factors. These
sources may originate from the measurement instrument, the environment,
the operator or the measured object itself (see Fig.~\ref{sources_of_uncertainty}). For example, uncertainty may arise from limited instrument resolution,
calibration imperfections, temperature variations, electronic noise or
surface irregularities of the measured component. In practice the total uncertainty associated with a measurement result
is the combined effect of all these contributions.

\section{Statistical interpretation}

Repeated measurements of the same quantity usually produce a set of
values that follow a statistical distribution. In many cases this
distribution is well approximated by a normal (Gaussian) distribution,
\begin{equation}
f(x)=\frac{1}{\sigma\sqrt{2\pi}}
\exp\left(-\frac{(x-\mu)^2}{2\sigma^2}\right),
\end{equation}

where $\mu$ represents the mean value and $\sigma$ represents the
standard deviation.

\begin{figure}[H]
\centering
\includegraphics[width=0.7\textwidth]{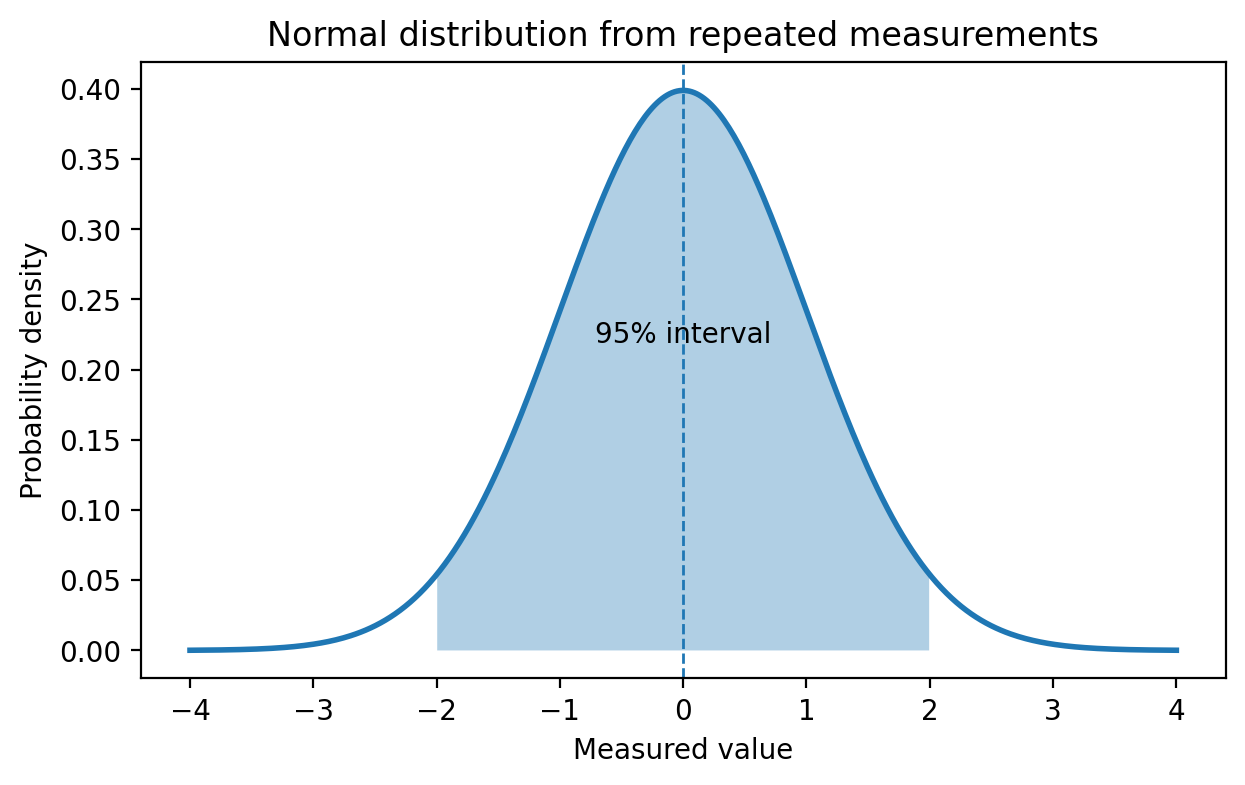}
\caption{Typical Gaussian distribution obtained from repeated
measurements.}
\end{figure}

The standard deviation provides a measure of the spread of measurement
results and is therefore closely related to measurement uncertainty.

\section{Confidence intervals}

Measurement uncertainty is often expressed using confidence intervals.

If the distribution of measurement results is approximately normal,
certain probabilities can be associated with intervals around the mean, for example
\begin{equation}
\mu \pm \sigma
\end{equation}

contains approximately 68\% of the measured values, while
\begin{equation}
\mu \pm 2\sigma
\end{equation}
contains approximately 95\% of the values.

The factor used to scale the uncertainty interval is known as the coverage factor and is usually denoted by $k$.

The expanded uncertainty is defined as

\begin{equation}
U = k u_c,
\end{equation}

where $u_c$ is the combined standard uncertainty.

In most engineering applications the value $k=2$ is used, corresponding
approximately to a 95\% confidence interval.

\section{Evaluation of uncertainty}

The internationally accepted framework for evaluating measurement
uncertainty is provided by the \textit{Guide to the Expression of
Uncertainty in Measurement (GUM)} \cite{GUM2008}.

According to this framework, uncertainty contributions are generally
classified into two categories.

\subsection{Type A evaluation}

Type A uncertainty is evaluated using statistical analysis of repeated
measurements. For example, if a~quantity is measured $n$ times, the mean value is
\begin{equation}
\bar{x} = \frac{1}{n}\sum_{i=1}^{n}x_i
\end{equation}

and the experimental standard deviation is

\begin{equation}
   s = \sqrt{\frac{1}{n-1}\sum_{i=1}^{n}(x_i-\bar{x})^2}. 
\end{equation}
The standard deviation provides an estimate of the uncertainty
associated with the measurement process.

\subsection{Type B evaluation}

Type B uncertainty is evaluated using information other than repeated
measurements. This may include calibration certificates, manufacturer
specifications, previous measurement data or expert judgement.

For example, if an instrument has a resolution of \SI{0.01}{mm}, the
uncertainty associated with that resolution can be estimated using an
appropriate probability distribution.

\section{Combined uncertainty}

When several uncertainty sources contribute to a measurement result,
their combined effect must be evaluated.

If the individual contributions are independent, the combined standard
uncertainty is obtained using the root-sum-of-squares formula

\begin{equation}
u_c = \sqrt{u_1^2 + u_2^2 + \cdots + u_n^2}.
\end{equation}

This equation is known as the law of propagation of uncertainty and
forms the basis of the GUM methodology.

\section{Importance of uncertainty evaluation}

Evaluating measurement uncertainty is essential for several reasons.

First, it allows measurement results to be compared objectively.
Second, it ensures that measurements are traceable to recognised
standards. Finally, it provides confidence in the reliability of
engineering and scientific measurements.

Without a rigorous evaluation of uncertainty, measurement results cannot
be interpreted correctly and may lead to incorrect decisions in
engineering design or quality control.

For these reasons, uncertainty analysis is a fundamental component of
modern metrology and measurement science.

\section{Estimating measurement uncertainty}

Evaluating measurement uncertainty is a structured process. In practical
applications, the goal is to identify all relevant sources of uncertainty
affecting a measurement and to combine their effects in order to estimate
the overall reliability of the measurement result.

The general framework for this process is described in the
\textit{Guide to the Expression of Uncertainty in Measurement (GUM)}
\cite{GUM2008}. Although the details depend on the measurement problem,
uncertainty evaluation typically follows a sequence of steps that guide
the analysis from the definition of the measurement to the reporting of
the final result.

\subsection{Eight steps for uncertainty evaluation}

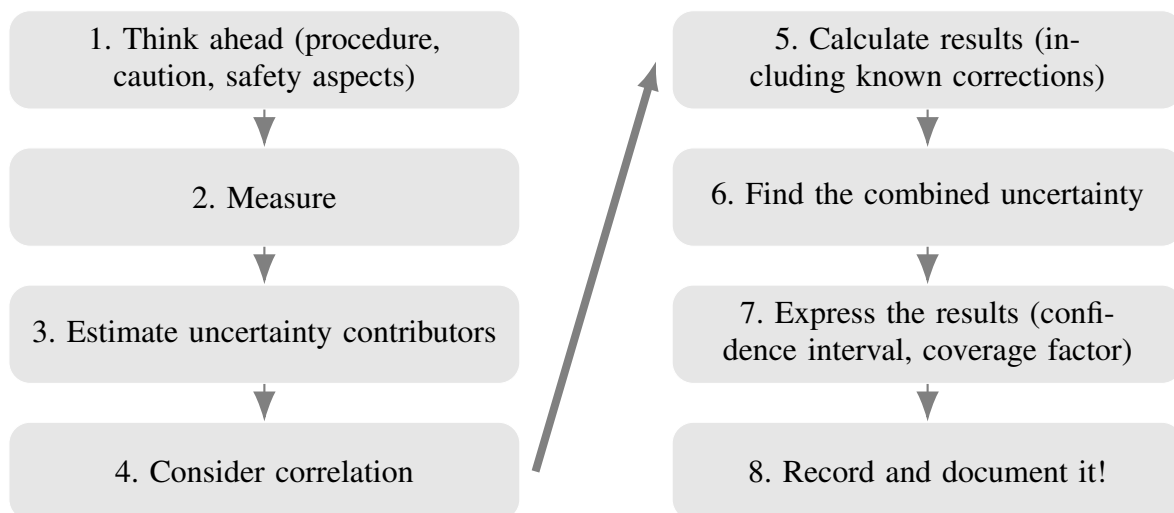
\begin{figure}[h]
\centering

\begin{tikzpicture}[
    node distance=.5 cm,
    box/.style={
        rectangle, 
        rounded corners=8pt, 
        minimum width=6cm, 
        minimum height=1.3cm, 
        text centered, 
        text width=6.5cm,
        draw=white,
        font=\large
    },
    arrow/.style={
        thick, 
        -{Latex[length=4mm]}, 
        draw=gray
    }
]

\node[box, fill=gray!20, text=black] (n1)
{1. Think ahead (procedure, caution, safety aspects)};

\node[box, fill=gray!20, text=black, below=of n1] (n2)
{2. Measure};

\node[box, fill=gray!20, text=black, below=of n2] (n3)
{3. Estimate uncertainty contributors};

\node[box, fill=gray!20, text=black, below=of n3] (n4)
{{4. Consider correlation}};

\node[box, fill=gray!20, text=black, right=2cm of n1] (n5)
{5. Calculate results (including known corrections)};

\node[box, fill=gray!20, text=black, below=of n5] (n6)
{6. Find the combined uncertainty};

\node[box, fill=gray!20, text=black, below=of n6] (n7)
{7. Express the results (confidence interval, coverage factor)};

\node[box, fill=gray!20, text=black, below=of n7] (n8)
{8. Record and document it!};

\draw[arrow] (n1) -- (n2);
\draw[arrow] (n2) -- (n3);
\draw[arrow] (n3) -- (n4);

\draw[arrow] (n5) -- (n6);
\draw[arrow] (n6) -- (n7);
\draw[arrow] (n7) -- (n8);

\draw[arrow, line width=3pt] 
([xshift=.2cm]n4.east) -- ([xshift=-0.2cm]n5.west);

\end{tikzpicture}

\caption{Measurement and uncertainty evaluation workflow.}
\label{seven_steps}
\end{figure}

The uncertainty evaluation process follows the general workflow introduced in Fig.~\ref{seven_steps}. In practical applications, the procedure begins by defining the measurand, selecting an appropriate measurement procedure and identifying the dominant uncertainty contributors associated with the instrument, environment, operator, workpiece and measurement method. Known systematic effects are corrected where possible, after which the remaining uncertainty contributions are propagated and combined to determine the overall uncertainty associated with the measurement result. The final stage consists of reporting the result together with the associated confidence interval and documenting the complete evaluation procedure to ensure traceability and reproducibility.

\subsection{Methods for estimating uncertainty}

Several approaches can be used to estimate measurement uncertainty. The
choice of method depends on the complexity of the measurement model and
the available information about the uncertainty sources.

Three methods are commonly used in engineering and industrial metrology.

The first is the \textit{GUM method}, which evaluates uncertainty through
analytical propagation of uncertainty using a mathematical measurement
model. This method is widely used when the relationship between input
quantities and the measured quantity is known.

A second approach is the \textit{spreadsheet method}, often referred to
as an \textit{uncertainty budget}. In this method, all significant
sources of uncertainty are organised in a table, and their contributions
are combined to determine the overall uncertainty of the measurement.

A third approach is the \textit{Monte Carlo method}. This method uses
numerical simulation to propagate uncertainties through the measurement
model and is particularly useful when the measurement relationship is
complex or nonlinear.

Each of these methods will be described in detail in the following
sections.

\section{The GUM method}

The most widely used framework for evaluating measurement uncertainty
is described in the \textit{Guide to the Expression of Uncertainty in
Measurement (GUM)} \cite{GUM2008}. The GUM method provides a systematic
approach for estimating uncertainty by propagating the uncertainties
associated with input quantities through a mathematical measurement
model.

The central idea of the method is that the quantity being measured,
referred to as the \textit{measurand}, is typically not measured
directly. Instead, it is determined from a number of input quantities
that are measured or known from other sources.

\subsection{Measurement model}

In the GUM framework, the relationship between the measurand and the
input quantities is described by a measurement model

\begin{equation}
Y = f(X_1, X_2, \ldots, X_N),
\end{equation}

where

\begin{itemize}
\item $Y$ is the measurand,
\item $X_1, X_2, \ldots, X_N$ are the input quantities,
\item $f(\cdot)$ represents the functional relationship between them.
\end{itemize}

Each input quantity is associated with a measurement uncertainty.
The objective of the GUM method is to determine how these uncertainties
propagate through the measurement model and influence the uncertainty
of the measurand.

\subsection{Linearisation of the measurement model}

In many practical situations the measurement model may be nonlinear.
To simplify the propagation of uncertainties, the GUM approach
approximates the function $f$ using a first-order Taylor expansion
around the expected values of the input quantities.

This linearisation is a first-order Taylor expansion about the nominal input vector
$\mathbf{X}_0=(X_{1,0},\ldots,X_{N,0})$:

\begin{equation}
Y \approx f(\mathbf{X}_0) + \sum_{i=1}^{N}
\left.\frac{\partial f}{\partial X_i}\right|_{\mathbf{X}_0}(X_i-X_{i,0}).
\end{equation}

The coefficients

\begin{equation}
c_i = \left.\frac{\partial f}{\partial X_i}\right|_{\mathbf{X}_0}
\end{equation}

are known as \textit{sensitivity coefficients}. These coefficients
describe how sensitive the measurand is to variations in each input
quantity.

\subsection{Propagation of uncertainty}

Using the linearised model, the combined standard uncertainty of the
measurand can be obtained from the uncertainties associated with the
input quantities.

If the input quantities are uncorrelated, the combined standard
uncertainty is given by

\begin{equation}
u_c(Y) =
\sqrt{
\sum_{i=1}^{N}
\left(
\frac{\partial f}{\partial X_i}
\right)^2
u^2(X_i)
}.
\end{equation}

This expression is commonly known as the \textit{law of propagation of
uncertainty}. It shows that the~overall uncertainty depends on both
the magnitude of the uncertainties of the input quantities and the~sensitivity of the measurand to each input variable.

If correlations exist between input quantities, additional covariance
terms must be included in the~calculation:

\begin{equation}
u_c^2(Y)=\sum_{i=1}^{N}c_i^2u^2(X_i)
+2\sum_{i<j}c_ic_j\,\mathrm{cov}(X_i,X_j).
\end{equation}

This form is important in practical metrology because some uncertainty
contributors are not independent. For example, temperature, fixturing,
calibration and alignment effects may influence several input quantities
simultaneously.

\subsection{Combined and expanded uncertainty}

The result of the uncertainty propagation is the
\textit{combined standard uncertainty}, denoted $u_c(Y)$.

To express the measurement result with a specified level of confidence,
the combined uncertainty is multiplied by a coverage factor $k$,
leading to the \textit{expanded uncertainty}

\begin{equation}
U = k\,u_c(Y).
\end{equation}

In many engineering applications the value $k = 2$ is commonly used,
corresponding approximately to a confidence level of 95\%.

The measurement result is therefore reported in the form

\begin{equation}
Y = y \pm U .
\end{equation}
\subsection{Example: calliper measurement}

As a simple example, consider the measurement of a length using a
calliper. 

\begin{figure}[H]
\centering

\begin{tikzpicture}[scale=1]

\draw[thick] (0,0) -- (10,0);

\foreach \x in {0,0.5,...,10}
    \draw (\x,0.15) -- (\x,-0.15);

\draw[fill=gray!30] (1,0) rectangle (1.3,1.2);

\draw[fill=gray!30] (7,0) rectangle (7.3,1.2);

\draw[fill=gray!15] (1.3,-0.4) rectangle (7,-0.8);
\node at (4.15,-1.1) {Measured object};

\draw[->, thick] (1.15,1.6) -- (1.15,1.2);
\node at (1.15,1.9) {$X_1$};

\draw[->, thick] (7.15,1.6) -- (7.15,1.2);
\node at (7.15,1.9) {$X_2$};

\draw[<->, thick] (1.3,1.0) -- (7,1.0);
\node at (4.15,1.3) {$Y = X_2 - X_1$};

\end{tikzpicture}

\caption{Illustration of a calliper measurement. The measured length
$Y$ is obtained from the difference between the~two scale readings
$X_1$ and $X_2$.}

\label{fig:calliper_diagram}

\end{figure}
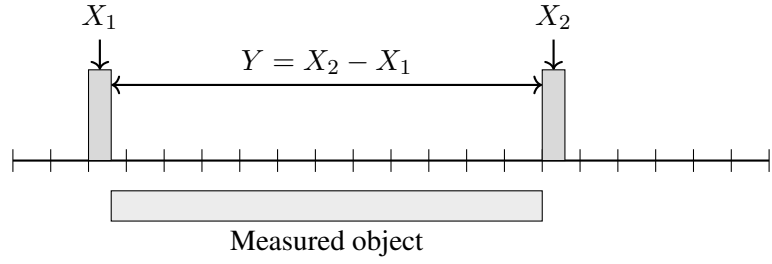

Suppose that the length $Y$ is obtained from two readings
$X_1$ and $X_2$ such that

\[
Y = X_2 - X_1 .
\]

Each reading is associated with an uncertainty of

\[
u(X_1) = u(X_2) = 0.1 \text{ mm}.
\]

Using the law of propagation of uncertainty, the combined uncertainty
is

\[
u(Y) =
\sqrt{
\left(\frac{\partial Y}{\partial X_1}\right)^2 u^2(X_1)
+
\left(\frac{\partial Y}{\partial X_2}\right)^2 u^2(X_2)
}.
\]

Since

\[
\frac{\partial Y}{\partial X_1} = -1
\qquad
\frac{\partial Y}{\partial X_2} = 1,
\]

the combined standard uncertainty becomes

\[
u(Y) =
\sqrt{(-1)^2 (0.1)^2 + (1)^2 (0.1)^2}
= \sqrt{0.01 + 0.01}
= 0.141 \text{ mm}.
\]

Assuming a coverage factor $k = 2$, the expanded uncertainty is

\[
U = 2 \times 0.141 = 0.282 \text{ mm}.
\]

If the measured length is $2.5$ mm, the final measurement result is
reported as

\[
Y = (2.50 \pm 0.28)\text{ mm}.
\]

This means that the true value of the length is expected to lie within
the interval

\[
2.22 \text{ mm} \le Y \le 2.78 \text{ mm}
\]

with a confidence level of approximately 95\%.
\subsection{Discussion}

The GUM method provides a rigorous and widely accepted framework for
uncertainty evaluation. It is particularly useful when the measurement
model is well defined and when the relationship between the~measurand
and the input quantities can be expressed analytically.

However, the method relies on the linearisation of the measurement
model and may become difficult to apply when the measurement
relationship is highly nonlinear or when the probability distributions
of the input quantities are not well known. In such cases alternative
approaches, such as Monte Carlo simulation, may be more appropriate.
\section{Spreadsheet method}

In practical engineering applications, the evaluation of measurement
uncertainty is often carried out using a structured tabular approach
known as the \textit{spreadsheet method} or \textit{uncertainty budget}.
This method provides a clear and systematic way of organising and
combining uncertainty contributions.

The spreadsheet method follows the principles of the GUM framework,
but presents the calculation in a form that is easier to implement in
practice, particularly when multiple uncertainty sources are involved.

\subsection{Uncertainty budget}

In this approach, all significant sources of uncertainty are identified
and listed in a table. Each contribution is characterised by its
estimated magnitude and by the probability distribution that best
describes its behaviour.

The purpose of the uncertainty budget is to ensure that all relevant
sources are taken into account and to provide a transparent overview of
the uncertainty evaluation process.

Each contribution is expressed as a \textit{standard uncertainty},
which corresponds to a standard deviation. For Type B contributions,
this is obtained by dividing the estimated uncertainty interval by a
factor that depends on the assumed probability distribution.

\begin{table}[H]
\centering
\fontsize{10}{12}\selectfont
\caption{Example of an uncertainty budget.}
\label{tab:uncertainty_budget}

\begin{tabular}{lcccc}
\hline \hline
\textbf{Source} & \textbf{Value} & \textbf{Distribution} & \textbf{Divisor} & \bm{$u_i$} \\
\midrule
Calibration & $\pm 0.03$ & Normal & 2 & 0.015 \\
Resolution  & $\pm 0.005$ & Rectangular & $\sqrt{3}$ & 0.0029 \\
Repeatability & --- & Normal & 1 & 0.010 \\
\hline \hline
\end{tabular}

\end{table}

\subsection{Combination of uncertainties}

Each uncertainty contribution is associated with a sensitivity
coefficient $c_i$, which describes how variations in the input quantity
affect the measurand. The combined standard uncertainty is obtained
using the~law of propagation of uncertainty in the form

\[
u_c = \sqrt{\sum_{i=1}^{N} (c_i u_i)^2}.
\]

Table~\ref{tab:uncertainty_contributions} shows the contribution of each
uncertainty source.

\begin{table}[H]
\centering
\fontsize{10}{12}\selectfont
\caption{Contribution of individual uncertainty sources.}
\label{tab:uncertainty_contributions}

\begin{tabular}{lcccc}
\hline \hline
\textbf{Source} & \bm{$u_i$} & \bm{$(c_i u_i)^2$} & \textbf{Contribution (\%)} \\
\midrule
Calibration & 0.015 & 0.000225 & 67.6 \\
Resolution  & 0.0029 & 0.000008 & 2.4 \\
Repeatability & 0.010 & 0.000100 & 30.0 \\
\midrule
\multicolumn{2}{r}{Total} & 0.000333 & 100 \\
\hline \hline
\end{tabular}

\end{table}

The combined standard uncertainty is therefore

\[
u_c = \sqrt{0.000333} = 0.018.
\]

\subsection{Expanded uncertainty}

To express the uncertainty with a specified level of confidence, the
combined standard uncertainty is multiplied by a coverage factor $k$:

\[
U = k\,u_c.
\]

For a coverage factor $k = 2$, the expanded uncertainty becomes

\[
U = 2 \times 0.018 = 0.036.
\]

The measurement result is then reported as

\[
Y = y \pm U.
\]

\subsection{Final result}

The final measurement result can be summarised as follows:

\begin{table}[H]
\centering
\fontsize{10}{12}\selectfont
\caption{Final measurement result.}
\label{tab:final_result}

\begin{tabular}{lc}
\hline \hline
Measured value & 50.00 mm \\
Combined uncertainty $u_c$ & 0.018 mm \\
Coverage factor $k$ & 2 \\
Expanded uncertainty $U$ & 0.036 mm \\
\hline \hline
\end{tabular}

\end{table}

\subsection{Modified spreadsheet example: calliper length measurement}

The same spreadsheet logic can be applied directly to the calliper model introduced in the GUM example. The measurand is

\begin{equation}
Y=X_2-X_1,
\end{equation}

where $X_1$ and $X_2$ are the two scale readings. The sensitivity coefficients are

\begin{equation}
c_1=\frac{\partial Y}{\partial X_1}=-1,\qquad
c_2=\frac{\partial Y}{\partial X_2}=1.
\end{equation}

Assuming independent standard uncertainties $u(X_1)=u(X_2)=\SI{0.10}{mm}$, the corresponding uncertainty budget is shown in Table~\ref{tab:caliper_budget}. This example is deliberately simple, but it is useful because it shows how the spreadsheet method reproduces the analytical GUM calculation while retaining a practical tabular format.

\begin{table}[H]
\centering
\fontsize{10}{12}\selectfont
\caption{Modified uncertainty budget for the calliper example.}
\label{tab:caliper_budget}
\begin{tabular}{lcccc}
\hline \hline
\textbf{Source} & \textbf{Standard uncertainty} & \textbf{Sensitivity} \bm{$c_i$} & \bm{$c_i u_i$} & \bm{$(c_i u_i)^2$} \\
\midrule
Reading $X_1$ & \SI{0.10}{mm} & $-1$ & $-\SI{0.10}{mm}$ & $0.0100$ \\
Reading $X_2$ & \SI{0.10}{mm} & $1$ & $\SI{0.10}{mm}$ & $0.0100$ \\
\midrule
\multicolumn{4}{r}{Sum of squared contributions} & $0.0200$ \\
\hline \hline
\end{tabular}
\end{table}

The combined standard uncertainty is therefore

\begin{equation}
u_c(Y)=\sqrt{0.0200}=\SI{0.141}{mm},
\end{equation}

and, using $k=2$, the expanded uncertainty is

\begin{equation}
U=2u_c(Y)=\SI{0.282}{mm}.
\end{equation}

For a measured length of $\SI{2.50}{mm}$, the result would be reported as

\begin{equation}
Y=(2.50\pm0.28)\,\si{mm}\qquad (k=2).
\end{equation}

\subsection{Interpretation of results}

The uncertainty budget allows the relative importance of each source
of uncertainty to be evaluated. In this example, the dominant
contribution arises from the calibration of the instrument, which
accounts for approximately 68\% of the total variance. The contribution
from repeatability is also significant, while the effect of the
instrument resolution is relatively small.

This type of analysis is one of the main advantages of the spreadsheet
method, as it allows the most significant sources of uncertainty to be
identified. In practice, this information can be used to improve the
measurement process by focusing efforts on reducing the dominant
contributions.

\subsection{Discussion}

The spreadsheet method provides a practical and transparent way of
evaluating measurement uncertainty and is widely used in industrial
metrology. It is particularly effective when the measurement model is
well understood and when uncertainty contributions can be clearly
identified and quantified.

However, the method relies on simplifying assumptions, such as the use
of linear uncertainty propagation and simple probability distributions.
For more complex or nonlinear measurement models, numerical methods such
as Monte Carlo simulation may be required.

\section{Monte Carlo method}

The Monte Carlo method provides a numerical approach for evaluating
measurement uncertainty by simulating the propagation of uncertainty
through a measurement model. Unlike the GUM method, which relies on
linearisation and analytical propagation, the Monte Carlo method uses
random sampling to obtain the distribution of the measurand directly.

This approach is particularly useful when the measurement model is
nonlinear, when input quantities have non-standard probability
distributions, or when correlations between variables are difficult to
handle analytically.

\subsection{Principle of the method}

In the Monte Carlo approach, each input quantity is treated as a random
variable described by a probability distribution. A large number of
random samples are generated for each input quantity, and the~measurement model is evaluated repeatedly using these sampled values.

If the measurement model is given by

\[
Y = f(X_1, X_2, \ldots, X_N),
\]

then each simulation produces one possible value of $Y$. Repeating this
process many times generates a distribution of output values, from which
the uncertainty can be determined.

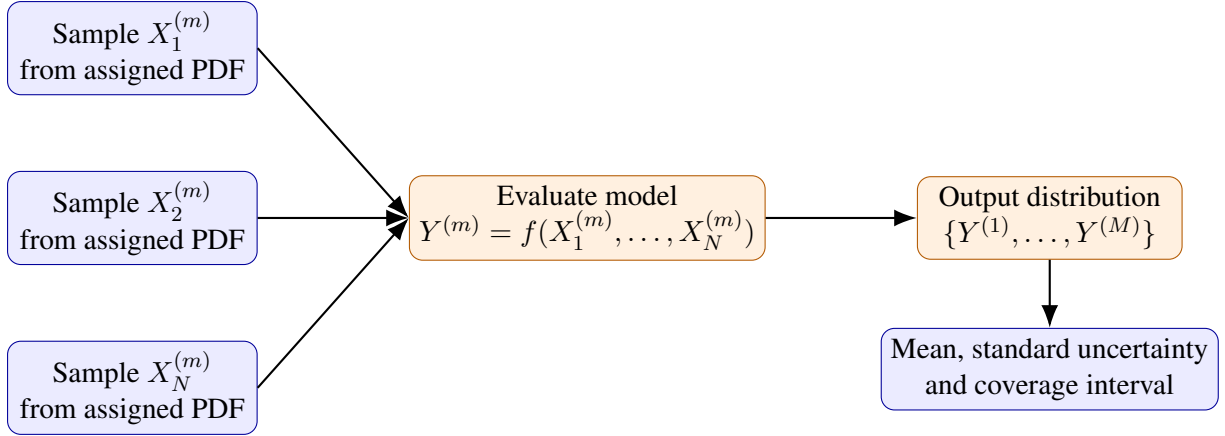
\begin{figure}[H]
\centering
\begin{tikzpicture}[
node distance=1.0cm,
box/.style={rectangle, rounded corners=5pt, draw=blue!60!black, fill=blue!8, minimum width=3.3cm, minimum height=0.9cm, align=center},
proc/.style={rectangle, rounded corners=5pt, draw=orange!70!black, fill=orange!12, minimum width=3.5cm, minimum height=1.0cm, align=center},
arrow/.style={-{Latex[length=3mm]}, thick}
]
\node[box] (x1) {Sample $X_1^{(m)}$\\from assigned PDF};
\node[box, below=of x1] (x2) {Sample $X_2^{(m)}$\\from assigned PDF};
\node[box, below=of x2] (xn) {Sample $X_N^{(m)}$\\from assigned PDF};
\node[proc, right=2.0cm of x2] (model) {Evaluate model\\$Y^{(m)}=f(X_1^{(m)},\ldots,X_N^{(m)})$};
\node[proc, right=2.0cm of model] (dist) {Output distribution\\$\{Y^{(1)},\ldots,Y^{(M)}\}$};
\node[box, below=0.9cm of dist] (stats) {Mean, standard uncertainty\\and coverage interval};
\draw[arrow] (x1.east) -- (model.west);
\draw[arrow] (x2.east) -- (model.west);
\draw[arrow] (xn.east) -- (model.west);
\draw[arrow] (model.east) -- (dist.west);
\draw[arrow] (dist.south) -- (stats.north);
\end{tikzpicture}
\caption{Monte Carlo propagation of uncertainty. Input quantities are sampled from their assigned probability distributions and propagated through the measurement model to obtain the output distribution of the measurand.}
\label{fig:mc_workflow}
\end{figure}

\subsection{Procedure}

The Monte Carlo method can be summarised as follows:

\begin{enumerate}
\item Define the measurement model $Y = f(X_1, \ldots, X_N)$,
\item Assign a probability distribution to each input quantity,
\item Generate random samples for each input quantity,
\item Evaluate the model for each set of sampled inputs,
\item Collect the resulting values of $Y$,
\item Analyse the distribution of $Y$ to determine uncertainty.
\end{enumerate}

\subsection{Estimation of uncertainty}

The output of the Monte Carlo simulation is a set of values

\[
\{Y_1, Y_2, \ldots, Y_M\},
\]

where $M$ is the number of simulations.

The mean value of the measurand is estimated as

\[
\bar{Y} = \frac{1}{M} \sum_{k=1}^{M} Y_k,
\]

and the standard uncertainty is obtained from the standard deviation of
the simulated values:

\[
u(Y) = \sqrt{
\frac{1}{M-1} \sum_{k=1}^{M} (Y_k - \bar{Y})^2
}.
\]

Confidence intervals can be determined directly from the distribution
of the simulated values, without relying on coverage factors or
assumptions of normality.

\subsection{Example: Monte Carlo evaluation of the calliper measurement}

The calliper measurement provides a useful example because the same model has already been treated analytically and by spreadsheet calculation. Consider again

\begin{equation}
Y=X_2-X_1.
\end{equation}

Let the measured readings be $x_1=\SI{0.00}{mm}$ and $x_2=\SI{2.50}{mm}$, with independent standard uncertainties

\begin{equation}
u(X_1)=u(X_2)=\SI{0.10}{mm}.
\end{equation}

In a Monte Carlo evaluation, pseudo-random samples are generated from assigned input distributions, for example

\begin{equation}
X_1^{(m)}\sim \mathcal{N}(x_1,u^2(X_1)),\qquad
X_2^{(m)}\sim \mathcal{N}(x_2,u^2(X_2)),
\end{equation}

and each simulated output is computed as

\begin{equation}
Y^{(m)}=X_2^{(m)}-X_1^{(m)},\qquad m=1,\ldots,M.
\end{equation}

The resulting sample $\{Y^{(m)}\}_{m=1}^{M}$ approximates the probability distribution of the measurand. For this linear model with Gaussian input distributions, the Monte Carlo output is also approximately Gaussian, with mean close to $\SI{2.50}{mm}$ and standard uncertainty close to

\begin{equation}
u(Y)=\sqrt{u^2(X_1)+u^2(X_2)}=\SI{0.141}{mm}.
\end{equation}

The corresponding expanded uncertainty for $k=2$ is therefore approximately $\SI{0.28}{mm}$, in agreement with the GUM and spreadsheet calculations.

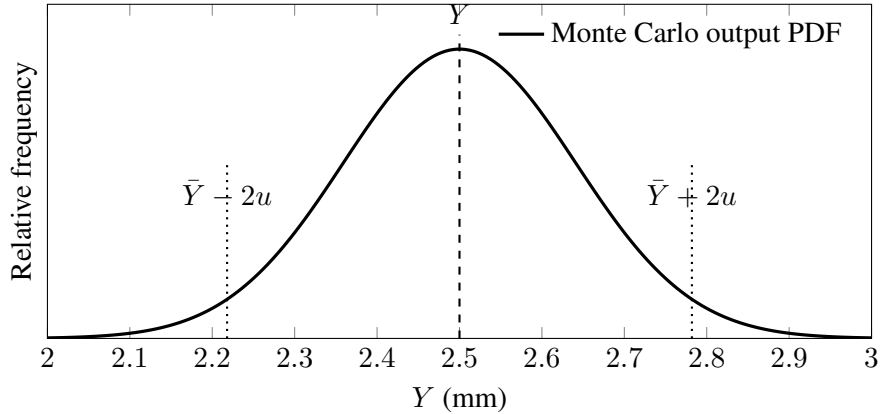
\begin{figure}[H]
\centering
\begin{tikzpicture}
\begin{axis}[
width=0.78\textwidth,
height=6cm,
xlabel={$Y$ (mm)},
ylabel={Relative frequency},
ytick=\empty,
xmin=2.0,xmax=3.0,
ymin=0,
legend style={draw=none, fill=none},
]
\addplot[very thick, domain=2.0:3.0, samples=200] {exp(-((x-2.5)^2)/(2*0.141^2))};
\addlegendentry{Monte Carlo output PDF}
\addplot[dashed, thick] coordinates {(2.5,0) (2.5,1.05)};
\addplot[dotted, thick] coordinates {(2.218,0) (2.218,0.6)};
\addplot[dotted, thick] coordinates {(2.782,0) (2.782,0.6)};
\node[anchor=south] at (axis cs:2.5,1.05) {$\bar{Y}$};
\node[anchor=north] at (axis cs:2.218,0.58) {$\bar{Y}-2u$};
\node[anchor=north] at (axis cs:2.782,0.58) {$\bar{Y}+2u$};
\end{axis}
\end{tikzpicture}
\caption{Monte Carlo interpretation of the calliper measurement. The simulated output distribution gives the mean value, standard uncertainty and coverage interval directly from the propagated samples.}
\label{fig:mc_caliper_output}
\end{figure}

\subsection{Graphical interpretation}

Figure~\ref{fig:mc_workflow} emphasises that Monte Carlo uncertainty evaluation is not a single algebraic calculation, but a~numerical propagation of input distributions through the measurement model. Figure~\ref{fig:mc_caliper_output} then shows how the output distribution can be summarised by its mean, standard uncertainty and coverage interval. This graphical interpretation is particularly valuable for nonlinear models, where the output distribution may be skewed, multimodal or otherwise poorly represented by a single symmetric interval.

\subsection{Advantages and limitations}

The Monte Carlo method has several advantages. It does not require
linearisation of the measurement model and can handle complex,
nonlinear relationships between variables. It also allows arbitrary
probability distributions to be used for input quantities and provides
a direct estimate of the output distribution.

However, the method requires a large number of simulations to obtain
accurate results and may therefore be computationally intensive.
Despite this, it is widely used in modern metrology, particularly in
applications where analytical methods are difficult to apply.

\subsection{Discussion}

The Monte Carlo method provides a flexible and powerful alternative to
analytical uncertainty evaluation. It is particularly useful when the
assumptions underlying the GUM method are not satisfied.

In practice, the choice between the GUM method, the spreadsheet method,
and the Monte Carlo method depends on the complexity of the measurement
model and the required level of accuracy.

\section{ISO 15530 methods for CMM measurement uncertainty}

In industrial dimensional metrology, uncertainty evaluation is often
performed using procedures defined in international standards. The ISO
15530 series provides practical methods for estimating task-specific
measurement uncertainty, particularly for coordinate measuring machines
(CMMs) and other coordinate measuring systems. These methods are important
because industrial coordinate measurements are affected simultaneously by
machine geometry, probing strategy, environmental effects, workpiece
properties, sampling strategy and evaluation software.

Unlike the GUM method, which relies primarily on analytical uncertainty
propagation, ISO 15530 includes experimental and simulation-based
approaches that aim to capture the combined influence of multiple
uncertainty sources under realistic measurement conditions. This makes the
standard particularly relevant when the complete functional relationship
between the measurand and all input quantities is difficult to write down
explicitly.

\subsection{ISO 15530-3: substitution method}

The substitution method estimates measurement uncertainty by measuring a
calibrated reference artefact that is similar to the workpiece and by using
the same measurement strategy, machine, environment and evaluation
algorithm. The essential idea is that the reference artefact samples the
same dominant uncertainty sources as the real measurement task.

In the first stage, a calibrated artefact is measured using the production
measurement procedure. The difference between the measured value and the
calibrated value provides an estimate of the task-specific measurement
error. In the second stage, the workpiece is measured under equivalent
conditions, and the uncertainty associated with the reference artefact and
measurement process is carried forward to the workpiece result.

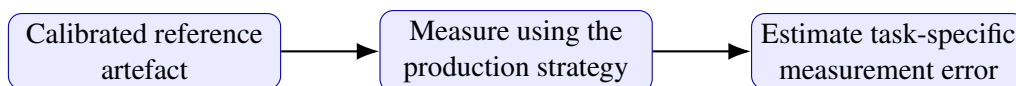
\begin{figure}[H]
\centering
\begin{tikzpicture}[node distance=1.0cm,
box/.style={rectangle, rounded corners=5pt, draw=blue!60!black, fill=blue!8, minimum width=3.6cm, minimum height=1.0cm, align=center},
arrow/.style={-{Latex[length=3mm]}, thick}]
\node[box] (a) {Calibrated reference\\artefact};
\node[box, right=1.3cm of a] (b) {Measure using the\\production strategy};
\node[box, right=1.3cm of b] (c) {Estimate task-specific\\measurement error};
\draw[arrow] (a) -- (b);
\draw[arrow] (b) -- (c);
\end{tikzpicture}
\caption{First stage of the ISO 15530-3 substitution method: a calibrated reference artefact is measured using the~same system and measurement strategy as the workpiece.}
\label{fig:iso_substitution_stage1}
\end{figure}

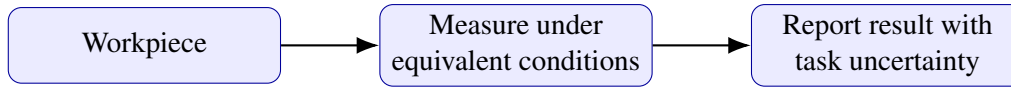
\begin{figure}[H]
\centering
\begin{tikzpicture}[node distance=1.0cm,
box/.style={rectangle, rounded corners=5pt, draw=blue!60!black, fill=blue!8, minimum width=3.6cm, minimum height=1.0cm, align=center},
arrow/.style={-{Latex[length=3mm]}, thick}]
\node[box] (a) {Workpiece};
\node[box, right=1.3cm of a] (b) {Measure under\\equivalent conditions};
\node[box, right=1.3cm of b] (c) {Report result with\\task uncertainty};
\draw[arrow] (a) -- (b);
\draw[arrow] (b) -- (c);
\end{tikzpicture}
\caption{Second stage of the ISO 15530-3 substitution method: the workpiece is measured under equivalent conditions and the task-specific uncertainty is assigned to the measurement result.}
\label{fig:iso_substitution_stage2}
\end{figure}

Because both measurements are performed under equivalent conditions, the
substitution method captures the combined influence of machine errors,
environmental effects, probing strategy, operator effects and evaluation
procedure. It is therefore simple, experimentally grounded and well suited
to repeated industrial measurement tasks. Its main limitation is that a
suitable calibrated artefact must be available and sufficiently similar to
the workpiece and measurand.

\subsection{ISO DTS 15530-4: Monte Carlo simulation}

ISO DTS 15530-4 extends task-specific uncertainty evaluation by introducing
numerical simulation of the coordinate measurement process. The method uses
a virtual representation of the measurement system, often referred to as a
Virtual Coordinate Measuring Machine (VCMM), to simulate uncertainty
contributors such as probing behaviour, geometric machine errors and
environmental effects.

\begin{figure}[H]
\centering
\includegraphics[width=0.95\textwidth]{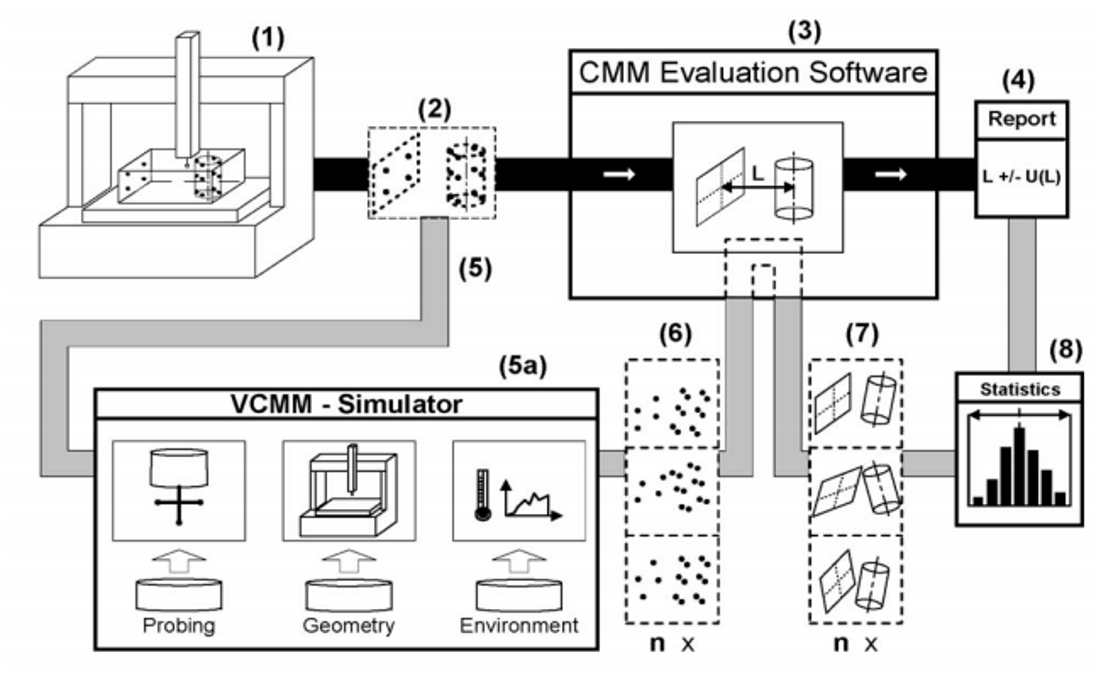}
\caption{ISO DTS 15530-4 Monte Carlo method. The method combines a real coordinate measurement with a virtual simulation of probing, geometry and environmental effects to estimate task-specific measurement uncertainty.}
\label{fig:iso15530_slide48}
\end{figure}

The motivation for the method is summarised in Fig.~\ref{fig:iso15530_slide48}.
For many practical CMM tasks, a useful uncertainty method must be flexible,
easy to apply, computationally efficient and require limited operator
intervention. The VCMM approach addresses this need by combining the real
measurement result with a simulation of the measurement process.

\subsubsection{First stage: real measurement}

\begin{figure}[H]
\centering
\includegraphics[width=0.95\textwidth]{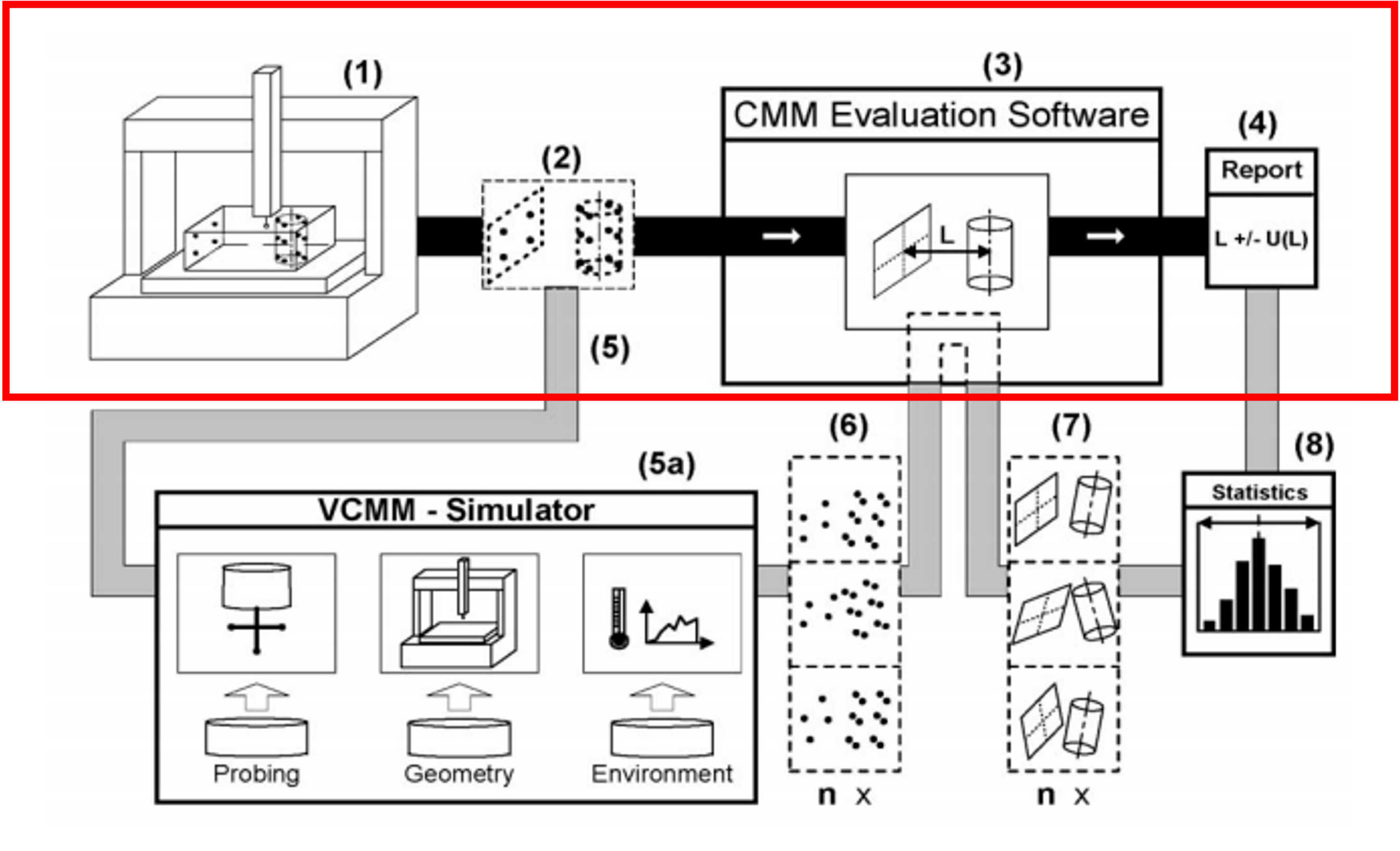}
\caption{First stage of the ISO DTS 15530-4 method: the coordinate measuring system performs the real measurement and produces the measurement result $Y$.}
\label{fig:iso15530_slide49}
\end{figure}

In the first stage, a normal measurement is carried out using the coordinate
measuring system, either contact or non-contact. Typical measurands include
flatness, perpendicularity, length and diameter. The~measured result $Y$ is
then taken as the starting point for the uncertainty simulation. This stage
corresponds to the upper part of the workflow in Fig.~\ref{fig:iso15530_slide49}.

\begin{figure}[H]
\centering
\includegraphics[width=0.95\textwidth]{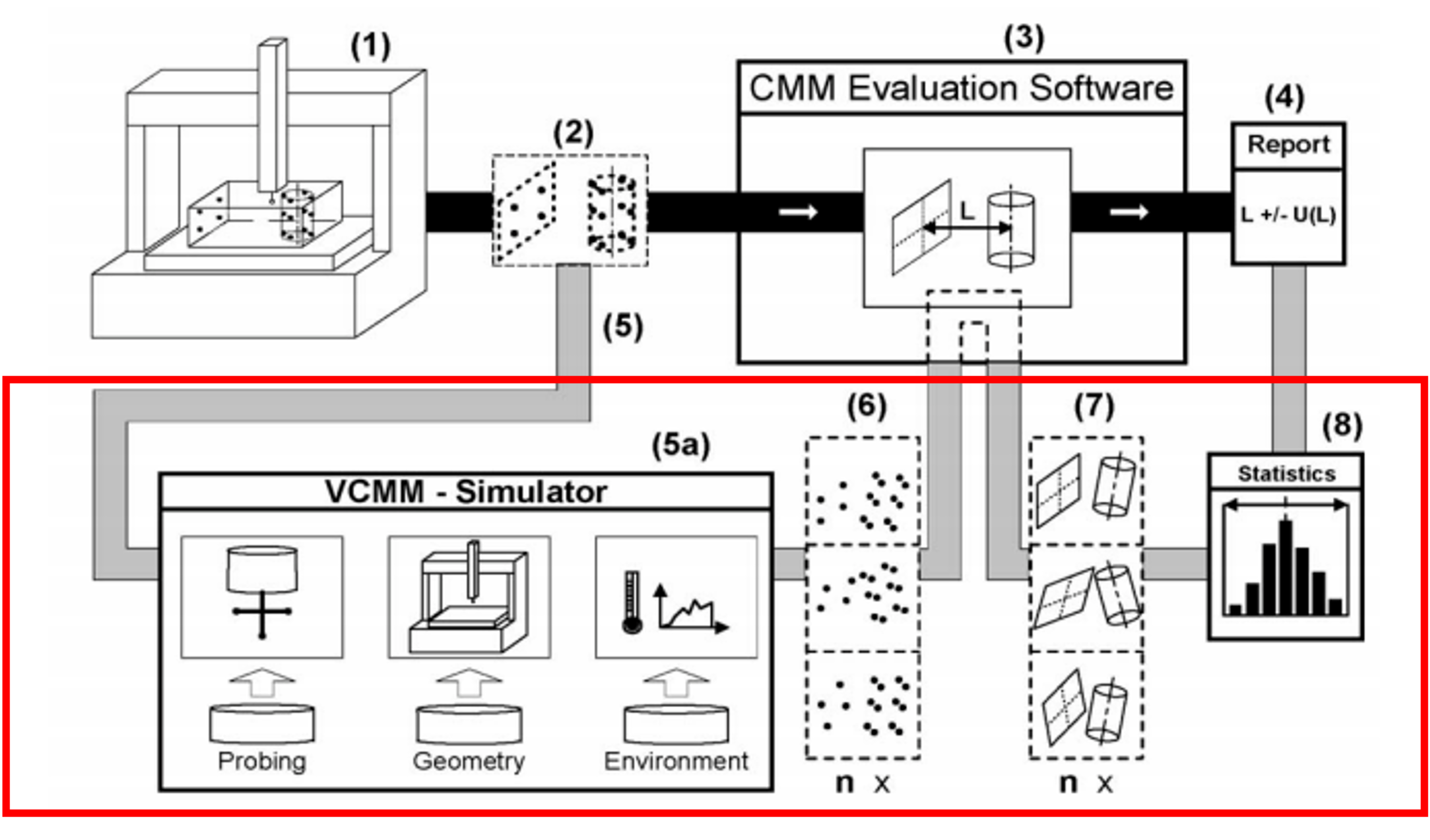}
\caption{Second stage of the ISO DTS 15530-4 method: measured points are perturbed according to uncertainty contributors and repeatedly processed with the same evaluation algorithm. The resulting simulated distribution is used to estimate uncertainty.}
\label{fig:iso15530_slide50}
\end{figure}

\subsubsection{Second stage: virtual perturbation and repeated simulation}

In the second stage, the measured points are perturbed according to the
identified uncertainty contributors and processed repeatedly using the same
evaluation algorithm. These perturbations represent the effects of probing,
machine geometry, environment and other contributors included in the VCMM.
After many repetitions, the distribution of simulated results is used to
estimate the standard uncertainty associated with the measurement task.

The combined standard uncertainty may be expressed as
\begin{equation}
u_c = \sqrt{u_{\mathrm{sim}}^2 + \sum_{i=1}^{n} u_i^2},
\end{equation}
where $u_{\mathrm{sim}}$ is the uncertainty contribution estimated by the
simulation and $u_i$ represents additional uncertainty sources not included
in the simulation.

\subsection{Discussion}

The ISO 15530 methods provide practical alternatives to purely analytical
uncertainty evaluation. The~substitution method is experimentally robust and
straightforward when suitable calibrated artefacts are available. The ISO
DTS 15530-4 Monte Carlo method is more flexible for complex measurement
systems because it uses simulation to represent the effects of machine,
probing, geometric and environmental contributors.

These approaches are particularly well suited to industrial applications,
where measurement systems are often too complex to model analytically. In
practice, they complement the GUM, spreadsheet and general Monte Carlo
methods by providing realistic estimates of measurement uncertainty under
actual operating conditions.

\section{Concluding remarks}

This chapter has introduced measurement uncertainty as the formal way of
attaching a reliable ``margin of doubt'' to a measurement result. A
measurement should not be interpreted as a single exact number, but as a
best estimate reported together with an uncertainty interval. Thus, a
result such as $L = (50.00 \pm 0.02)\,\mathrm{mm}$, means that \(50.00\,\mathrm{mm}\) is the best estimate of the length, while
the uncertainty expresses the~range of values that could reasonably be
attributed to the measurand.

A central message of the chapter is that error and uncertainty are not
the same. Error is the difference between the measured value and the true
value, but the true value is usually unknown. Uncertainty instead
quantifies how large the possible error may be, based on the available
information. Students should also distinguish between systematic and
random effects: systematic errors introduce bias and mainly affect
accuracy, while random errors introduce scatter and mainly affect
precision.

The statistical interpretation of uncertainty is essential. Repeated
measurements give information about variability, often represented by a
probability distribution. The standard deviation gives a measure of the
spread of the data, while the coverage factor \(k\) is used to convert a
combined standard uncertainty into an expanded uncertainty, $U = k u_c$. In many engineering applications, \(k=2\) is used to represent an~approximately \(95\%\) confidence interval.

The chapter has presented three main approaches for evaluating
measurement uncertainty. The~GUM method provides the analytical
framework, propagating input uncertainties through a measurement model.
The spreadsheet method expresses the same idea in a practical uncertainty
budget, allowing the dominant contributors to be identified. The Monte
Carlo method uses repeated numerical sampling and is especially useful
when the measurement model is nonlinear or when input distributions are
complex.

The ISO 15530 methods show how uncertainty evaluation is implemented in
industrial dimensional metrology, especially for coordinate measuring
systems. These methods are important because they account for realistic
measurement conditions, including machine behaviour, environment,
operator effects and task-specific measurement strategy.

The most important lesson is that uncertainty is not an
optional addition to a measurement. It is what makes the measurement
scientifically meaningful, traceable and useful for decision-making. A~measurement result without an uncertainty statement is incomplete,
because it does not indicate how much confidence should be placed in the
reported value.

\end{document}